\documentclass[12pt]{article}
\usepackage{epsfig,axodraw}
\input epsf.sty
\input axodraw.sty
\newcommand{\la}[1]{\label{#1}}
 
\newcommand{\be}{\begin{equation}}
\newcommand{\ee}{\end{equation}}
\newcommand{\ba}{\begin{eqnarray}}
\newcommand{\ea}{\end{eqnarray}}
\newcommand{\bi}{\begin{itemize}}
\newcommand{\ei}{\end{itemize}}

\newcommand{\tr}{{\rm Tr\,}}
\newcommand{\ii}{{\rm i}}
\newcommand{\trq}{{\rm \hat{T}r\,}}
\newcommand{\ex}{{\rm e}}
\newcommand{\re}{\mathop{\rm Re}}

\newcommand{\im}{\mathop{\rm Im}}
\newcommand{\nn}{\nonumber}

\newcommand{\bfx}{{\bf x}}
\newcommand{\bfy}{{\bf y}}
\newcommand{\bfz}{{\bf z}}

\renewcommand{\vec}[1]{{\bf #1}}

\newcommand{\h}{\cal{H}}
\newcommand{\hi}{\hat{\i}}

\newcommand{\hp}{\hat{P}}
\newcommand{\htm}{\hat{T}}
\newcommand{\op}{{\cal O}}

\newcommand{\RR}{{\rm I\kern -.2em  R}} 
\newcommand{\eq}{Eq.~}
\newcommand{\eqs}{Eqs.~}
\newcommand{\fig}{Fig.~}

\def\lsi{\raise0.3ex\hbox{$<$\kern-0.75em\raise-1.1ex\hbox{$\sim$}}}
\def\gsi{\raise0.3ex\hbox{$>$\kern-0.75em\raise-1.1ex\hbox{$\sim$}}}

\newcommand{\gsim}{\mathop{\gsi}}
\def\none               {\multicolumn{2}{c}{---}}
\def\noner               {\multicolumn{2}{c|}{---}}

\makeatletter \@addtoreset{equation}{section} \makeatother

\makeatletter
\renewcommand\section{\@startsection {section}{1}{\z@}%
                                   {-5.5ex \@plus -1ex \@minus -.2ex}% bfr-skip
                                   {2.3ex \@plus.2ex}%
                                   {\normalfont\large\bfseries}}
\renewcommand\subsection{\@startsection{subsection}{2}{\z@}%
                                     {-3.25ex\@plus -1ex \@minus -.2ex}%
                                     {1.5ex \@plus .2ex}%
                                     {\normalfont\normalsize\bfseries}}
\renewcommand\thesection {\@arabic\c@section}
\renewcommand\thesubsection   {\thesection.\@arabic\c@subsection}
\renewcommand{\@seccntformat}[1]{%
\csname the#1\endcsname.\hspace{1.0em}}
\makeatother
\begin{document}
 
\begin{titlepage}
\begin{flushright}
MIT-CTP-3217
\end{flushright}
\begin{centering}
\vfill
 
{\bf ON THE NON-PERTURBATIVE GLUON MASS \\ AND HEAVY QUARK PHYSICS}

\vspace{0.8cm}
 
Owe Philipsen\footnote{email: philipse@lns.mit.edu}

\vspace{0.3cm}
{\em 
Center for Theoretical Physics,\\ Massachusetts Institute of Technology,\\
Cambridge, MA 02139-4307, USA\\}

\vspace*{0.7cm}
 
\begin{abstract}
\noindent
We study a recently proposed gauge invariant, 
non-local pure gauge observable, which is equivalent to the gluon propagator
in a certain gauge. The correlator describes a gluon coupled to
static sources and decays with 
eigenvalues of the Hamiltonian, permitting a non-perturbative definition
of a gluonic parton mass. 
Detailed numerical tests of the observable are performed
in SU(2) gauge theories in 2+1 dimensions. In a Higgs model it reproduces
the physical W-boson mass, while in a confining theory
its non-local nature results in 
almost exclusive projection onto torelonic states.
However, the gluon mass can also be related to the
mass difference between the lowest gluelumps, 
i.e. gluon configurations
bound to adjoint sources. Its value is computed in SU(2) pure gauge theory
in 2+1 dimensions to be $m_A=0.360(19)g^2$, which plays an important role as 
``magnetic mass" of the four-dimensional theory at high temperatures.
The same quantity is found to determine the mass splitting between vector and 
scalar mesons in the three-dimensional theory.
In SU(3) pure gauge theory in 3+1 dimensions 
the corresponding gluelump splitting yields
$m_A\approx 370$ MeV. Possible relations of this quantity with the 
QCD heavy meson spectrum are discussed.
\end{abstract}
\end{centering}
%
%
 
%\noindent
%PACS numbers: 
%12.38.Mh, %        Quark-gluon plasma
%11.10.Wx, %        Finite temperature field theory
%12.38.Gc, %       Lattice QCD calculations
%11.10.Kk, %        Field theory in dimensions other than four.
%11.15.Pg, %        Expansions for large numbers of components 
           %        (e.g., 1/N sub c expansions)
%24.85.+p  %        Quarks, gluons, and QCD in nuclei and nuclear processes
%\\
%Keywords:
%QCD,
%finite temperature,
%finite density,
%dimensional reduction,
%quark gluon plasma,
%Debye mass.

\vfill

\end{titlepage}
 
%%%%%%%%%%%%%%%%%%%%%%%%%%%%%%%%%%%%%%%%%%%%%%%%%%%%%%%%%%%%%%%%%%%%%%%%%%
%%%%%%%%%%%%%%%%%%%%%%%%%%%%%%%%%%%%%%%%%%%%%%%%%%%%%%%%%%%%%%%%%%%%%%%%%%

\section{Introduction}

The physics of confinement of quarks in QCD is well established experimentally
and by numerical simulations of lattice gauge theories. 
However, little is known about
the confiment mechanism, i.e.~how parton physics at short 
distances smoothly evolves into hadron physics at large distances.
Understanding this transition requires non-perturbative
knowledge of the parton dynamics at all length scales. 
Such knowledge is indispensable in the
context of finite temperature and baryon density, where matter is believed
to appear in a deconfined state whose
collective physical properties 
are determined by parton dynamics.
The latter is encoded in the Green functions of quark
and gluon fields, which 
in general are not gauge invariant. This raises the conceptual
problem of how to define non-perturbative partons and extract 
physical information about them.
In this work an attempt is made to answer this question for a suitably defined 
gluon correlation function. 
Once this most elementary case is understood, it may be generalized to
quarks and more complex Green functions.

In perturbation theory one fixes a gauge and studies partons directly.
Although field propagators are gauge dependent and not physical observables,
they carry gauge invariant 
information about the parton dynamics in their singularity structure.
Pole masses defined from the gluon and quark propagators have been proved to be gauge
independent to every finite order in perturbation theory \cite{kkr,kro}.
However, perturbation theory is limited by the requirement
of weak coupling. On the other hand, numerical results obtained by fixing a gauge
on the lattice \cite{mo} have often been inconclusive or 
controversial in the past. This is due to the difficulty 
to fix a gauge uniquely and avoid
the problem of Gribov copies \cite{grib}.
Moreover, most complete gauge fixings (e.g. the Landau gauge)
violate the positivity of the transfer matrix, thus obstructing
a quantum mechanical interpretation of the results.
Finally, gauge independence of any result is not evident, but has to be
demonstrated numerically by comparing different gauges, while it is
extremely difficult to control the mentioned 
problems for each gauge.
An overview with references to numerical work
may be found in \cite{rev}.
 
A non-perturbative argument for a non-zero 
dynamical mass scale associated
with gluons was given a long time ago \cite{ber}, based on the
static potential of adjoint sources. If gluons were non-perturbatively massless,
they could be pair produced at no energy cost and the adjoint potential would
be screened even for small separations. 
In lattice simulations, by contrast, one observes a
linear rise up to rather large distances, until the adjoint string
breaks at $\sim 10 M_G^{-1}$, in units of the lightest glueball \cite{adj1,adj2}.

Recently, non-local, gauge invariant observables have been constructed in 
lattice gauge theory which
are equivalent to the gluon propagator in certain gauges. 
The transfer matrix formalism has been used to prove 
that these gluon correlators decay exponentially
with eigenvalues of the Hamiltonian \cite{me},
implying a pole structure in momentum space.
The energy gap between the vacuum and the lowest excited eigenvalue then
represents a gauge invariant definition for a gluonic parton mass.

The purpose of the current work is to compute the value of the gluon mass
in pure gauge theory and to establish its relation to physical observables. 
The main results are as follows:
A gauge invariant probe of the gluon energy is obtained by
coupling it to external static sources and 
measuring the energy of the composite object. The problem is 
then reduced to separate the contribution of the gluon field from that of
the sources.  In principle, this can be achieved in two ways. First, by dressing the 
field variables with a non-local functional representing an 
average over all positions of the
sources. Second, by introducing explicit fields for the sources, 
which can be integrated over analytically.
In both cases the resulting gauge invariant observables can be related to 
the gluon propagator in a particular gauge. 
The second procedure has the advantage of being local and numerically feasible 
in a confining theory. 
Moreover, it lends itself to a straightforward 
interpretation of the gluon mass, whose value equals that of the
level splitting between the lightest ``gluelumps'', i.e.~gluonic configurations 
coupled to an adjoint source.
The latter can be computed straightforwardly without need 
for gauge fixing or introduction of non-local fields.
We calculate its value for SU(2) Yang-Mills theory in 2+1 dimensions and find 
$m_A=0.36(2)g^2$. This result plays an important role for the four-dimensional 
high temperature physics,
where it corresponds to the magnetic mass with $g^2\sim g_{4d}^2T$.
For SU(3) in 3+1 dimensions, the same splitting has been computed in the literature
to be $m_A=368(7)$ MeV \cite{glmic}. 
In three dimensions one furthermore finds 
the splitting of heavy vector and scalar
mesons to be given by the gluon mass.
The situation in QCD might be similar, but a careful study of the static limit
is necessary to settle this question.

The outline of the paper is as follows.
In Sec.~\ref{ham} we review the Hamiltonian formalism of lattice gauge theory,
which in its strong coupling limit permits a natural definition 
of a gauge invariant mass gap in terms of a gluon coupled to external sources.
Sec.~\ref{nlgl} recalls the main results of \cite{me}, relating the Hilbert space 
definition to a non-local Euclidean observable 
amenable to numerical simulations. This operator is tested in
2+1 dimensional theories in Sec.~\ref{num}.
The relation to the gluelump
spectrum is derived in Sec.~\ref{stati}, and numerical results are quoted for
SU(2) in 2+1 and SU(3) in 3+1 dimensions. Finally, Sec.~\ref{qcd} discusses the connection
to the heavy meson spectrum for SU(2) in 2+1 dimensions as well as QCD,
before conclusions are given in Sec.~\ref{con}.
A continuum formulation of gluons probed by sources as well as 
lattice actions and parameters used in the simulations are 
given in two appendices.

\section{\label{ham} The gluon in the Hamiltonian formalism}

To begin, it is instructive to recall the Hamiltonian formulation of lattice gauge
theory \cite{ks}, where it is easy to see that 
gauge invariant information is associated with field variables and how it can be interpreted
physically. We are interested in SU(N) 
gauge theory on an $L^d\times N_t$ lattice with periodic
boundary conditions.
In a Hilbert space formulation \cite{ks,cre,pos} a
spatial sublattice $L^d$ at a fixed time is considered, with
link variables $U_i(\bfx)$ and the corresponding field operators $\hat{U}_i(\bfx)$.
The wave functions form a Hilbert space ${\h}_0$ of all complex,
square integrable functions
$\psi[U]$ defined on the gauge group $G$:
${\h}_0=[L^2(G)]^{d L^d}$.
A unitary operator $\hat{R}(g)$ imposes gauge transformations
$g(\bfx)\in G$ on wave funtions according
to
\be \label{gt}
\left(\hat{R}\psi\right)[U] =\psi[U^g], 
\qquad U_i^g(\bfx)=g(\bfx)U_i(\bfx)g^{\dag}(\bfx+\hi)\,.
\ee
Wave functions of physical particle states are gauge invariant, $\psi[U^g]=\psi[U]$,
thus forming a subspace $\h\subset$${\cal H}_0$. 
Any wave function $\psi[U]\in {\h}_0$ can be projected on the physical subspace by means
of the projection operator $\hp$,
\be
(\hp \psi)[U]=\int \prod_{\bfx} dg(\bfx)\;\psi[U^g].
\ee

The Kogut-Susskind Hamiltonian acting on the full space ${\cal H}_0$ 
is obtained by quantizing
the theory in temporal gauge
$U_0=1$ and reads \cite{ks}
\be \label{hks}
\hat{H}_0
=\frac{g^2}{2a^{d-2}}\sum_{\bfx,i} (\hat{E}_i^a(\bfx))^2 -\frac{2}{g^2a^{4-d}}
\sum_{\bfx,i<j}\re\tr\hat{P}_{ij}(\bfx)\,,
\ee
where $P_{ij}(\bfx)=U_i(\bfx)U_j(\bfx+\hat{\i})U_i^\dag(\bfx+\hat{\i}+\hat{\j})U_j^\dag(\bfx)$
is the elementary plaquette field.
The components of electric field and link operators 
satisfy the commutation relations
\ba
\left[\hat{E}_i^a(\bfx),\hat{E}_j^b(\bfy)\right]&=&
\ii f_{abc}\hat{E}_i^c(\bfx)\delta_{ij}\delta_{\bfx\bfy}, \nn \\
\left[\hat{E}_i^a(\bfx),\hat{U}_{j,\alpha\beta}(\bfy)\right] &=& 
-T^a\hat{U}_{i,\alpha\beta}(\bfx)\delta_{ij}\delta_{\bfx\bfy}.
\ea
In principle the whole configuration space may be constructed 
by applying link operators in all possible combinations to the vacuum state, $\psi_0=1$. 
In particular, doing this just once we have
\be
%\hat{E}_i^a(\bfx) \psi_0=0,\quad 
\psi^1_{\alpha\beta}[U]\equiv
{\hat{U}}_{i,\alpha\beta}(\bfx)\psi_0=
U_{i,\alpha\beta}(\bfx)\psi_0.
\ee
Thus, $\psi^1[U]$ is the wave function corresponding to one link variable being excited
and all others being in the vacuum configuration.
Clearly, it transforms non-trivially,
$\psi^1_{\alpha\beta}[U^g]=
g_{\alpha\gamma}(\bfx)g^*_{\beta\delta}(\bfx+\hat{\i})\psi_{\gamma\delta}[U]$. 
Gauge invariant wave functions are obtained by exciting closed loops of links, the simplest
being a plaquette,
$\psi^P[U]=P_{ij}(\bfx)$.

In the limit of strong coupling, $a^{(3-d)}g^2 \rightarrow \infty$, the potential term
in the Hamiltonian decouples. Although this limit is far from the continuum and 
the physical situation, it illustrates the structure of the
Hilbert space ${\cal H}_0$. 
In the strong coupling limit $\psi^1$ is an eigenstate of the Hamiltonian,
\be \label{strong}
\hat{H}_0\psi_{\alpha\beta}^1[U]=E_1\psi_{\alpha\beta}^1[U],
\quad E_1=\frac{g^2}{2a^{d-2}}C_F, \quad C_F=(N^2-1)/2N.
\ee
Excitation of other links increases the energy by one unit for every link.
The wave function for a string of electric flux between charges at $\bfx, 
\bfy=\bfx+l\hat{\i}$,
\be \label{psis}
\psi^S_{\alpha\beta}[U]=[U_i(\bfx)U_i(\bfx+\hat{\i})\ldots 
U_i(\bfx+(l-1)\hat{\i})]_{\alpha\beta}=U_{\alpha\beta}(\bfx,\bfy),
\ee
is an eigenstate with energy $lE_1$, whereas $\psi^P[U]$ has energy $4E_1$.
The plaquette wave function is in the Hilbert space of gauge invariant functions,
$\psi^P[U]\in \h$.
On the other hand, the wave functions $\psi^{1,S}$ 
are non-invariant and hence not elements
of the projected Hilbert space. 
Nevertheless, because time independent gauge transformations commute with the Hamiltonian, 
$[\hat{R},\hat{H}]=0$, the eigenvalues belonging to these wave functions 
are gauge invariant. 

It is now natural to define a ``gluon mass gap'' by the energy it takes 
to excite one gluon configuration out of the vacuum.
Defining the usual lattice gluon field
\be
A_i(\bfx)=\frac{\ii}{2g}
\left[U_i(\bfx) -U^{\dag}_i(\bfx)
-\frac{1}{N}\tr\left(U_i(\bfx) -U^{\dag}_i(\bfx)\right)\right],
\ee
one has for the corresponding wave function 
$\psi^A_{\alpha\beta}[U]=\hat{A}_{i,\alpha\beta}(\bfx)\psi_0$ in the strong coupling limit
\be \label{gluestrong}
\hat{H}_0\psi^A_{\alpha\beta}[U]=E_1\psi^A_{\alpha\beta}[U].
\ee
Hence, in the Hamiltonian strong coupling limit the gluon has a gauge invariant mass gap,
which is an eigenvalue of $\hat{H}_0$ to a gauge covariant eigenfunction. 
Note that the wave functions for a unit of flux and the gluon wave function 
are constructed from the same link variable. 
The difference is in the transformation property. $\psi^A$ picks the traceless
part of the link variable only, and hence is an ($N^2-1$)-plet
of the form $\sim c^a[U]T^a$ with
some real functions $c^a[U]$. 

Away from the strong coupling limit the energy levels and the precise form of the 
eigenstates will change. 
However, their transformation properties remain the same 
since gauge transformations commute with the full Hamiltonian.
Quite generally, eigenstates with non-trivial transformation behaviour 
describe strings of flux with energy $E$.
They can be made manifestly gauge invariant and ``physical'' by letting the
flux end on static sources. The total wave functions then are gauge invariant combinations
of the wave function for the source, $\phi^g(\bfx)=g(\bfx)\phi(\bfx)$,
and the gauge part,
\be \label{psitot}
\psi^{1,S}_{tot}[U]=\phi^*_\alpha(\bfx)\phi_\beta(\bfx+\hat{\i}) \psi^{1,S}_{\alpha\beta}[U], 
\quad
\psi^A_{tot}[U]=\frac{\ii}{2g}\left(\psi^{1}_{tot}[U]-\psi^{1\dag}_{tot}[U]\right).
\ee
 
In summary, in addition to the gauge invariant particle states of $\h$, 
${\cal H}_0$ contains additional 
states describing
static potentials, or gauge field configurations probed by external charges. 
The corresponding pure gauge wave
functions transform non-trivially, but the associated energies are fully
gauge invariant and contain valuable information about the dynamics of charged
states. In this respect a gluon wave function is on the same
footing as a wave function for a flux element, whose traceless part
it constitutes.
We conclude that a gluon mass
may be defined by the minimal energy it takes to excite a flux element in 
an $(N^2-1)$ or adjoint state.
This energy is {\it not} related to an asymptotic particle state,
but rather a field coupled to static sources. The expression \eq (\ref{psitot})
suggests that it is a short distance 
property of the static potential with the sources 
and the flux in an adjoint state.

\subsection{Transfer matrix and Euclidean correlation functions}

In order to make the spectrum amenable
to numerical simulations, the Hilbert space picture has to be connected to
Euclidean correlation functions 
by means of the transfer matrix \cite{cre,pos}. 
The latter is an integral operator translating the wave functions in time,
\ba
\psi[U_{t+1}]&=&\left(\htm_0 \psi\right)[U_t]\nn\\
&=&\int\prod_{\bfx,i}dU_{i,t}(\bfx)\;\exp -S_t[U_{t+1},U_t] \psi[U_t],
\ea
where $S_t$ is the Wilson action of two neighbouring timeslices in temporal gauge.
$\htm_0$ is a bounded, self-adjoint operator
with a strictly positive spectrum \cite{pos},
allowing to define a bounded Hamiltonian $\hat{H}_0=- 1/a\ln \htm_0$,
which up to corrections $\op(a^2)$, is identical to the Kogut-Susskind Hamiltonian
\eq (\ref{hks}).
Through the projection 
\be
\htm=\hp\htm_0
\ee
one defines a transfer matrix $\htm$ acting on the gauge invariant subspace only,
as well as the corresponding Hamiltonian $\hat{H}$.

Writing the quantum mechanical trace over a complete set of states on the space
$\h$ as
$
\trq{\op}=\sum_n\langle n|\op| n \rangle
$, a one to one
relation between Euclidean expectation values and quantum mechanical 
traces can be established.
For correlators of gauge invariant, local operators $\op(x)$
the result is
\ba
\langle \op(\bfx,0) \op(\bfx,t)\rangle &=& 
Z^{-1}\trq\{\htm^{N-t}\hat{\op}(\bfx)\htm^t\hat{\op}(\bfx)\}\nn\\
&\stackrel{N_t\rightarrow \infty}{\longrightarrow}&
\sum_n
|\langle 0| \hat{\op}(\bfx)|n\rangle |^2 \exp -(E_n-E_0)t\;.
\ea
The eigenvalues and states $|n>$ are those of the Hamiltonian $\hat{H}$
and describe the asymptotic particle states, in the pure gauge theory the glueballs. 

On the other hand, the static potential is probed by the Wilson loop, with
the temporal lines representing the external charges. 
In this case the correspondence is
\ba \label{qmpot}
\langle W(|\bfx-\bfy|,t)\rangle &=&  
Z^{-1}\trq\{\htm^{N_t-t}\hat{U}_{\alpha\beta}(\bfx,\bfy)\; \htm_0^t \; 
\hat{U}^{\dag}_{\alpha\beta}(\bfx,\bfy)\}\nn\\
&\stackrel{N_t\rightarrow\infty}{\longrightarrow}& 
\sum_n
|\langle 0| \hat{U}_{\alpha\beta}(\bfx,\bfy)|n\rangle |^2 \exp -(E_n-E_0)t\;.
\ea
Here the $|n>$ and $E_n$ are eigenstates and eigenvalues of 
$\hat{H}_0$ and do {\it not} correspond to particle states, but rather to
field energies in the presence of external charges.
Quite generally, correlators involving temporal Wilson lines describe
configurations involving static charges, their expontential fall-off is
dictated by $\hat{H}_0$, and the intermediate states transform non-trivially.

Euclidean and Hilbert space formalism are but a different 
language for the same physics. Following the observations in the last section, 
it must then be possible to construct a Euclidean operator $O[U]$ whose
expectation value can be expressed analogous to \eq (\ref{qmpot})
with $\hat{U}(\bfx,\bfy)\rightarrow \hat{A}_i(\bfx)$.

\section{\label{nlgl} A non-local, gauge invariant gluon operator}

It has recently been shown that such a Euclidean expression can be realized with
an auxiliary complex $N$-plet $f[U]$  
transforming in the fundamental
representation of the group \cite{me}.
This procedure is not unique. One possibility is to use
eigenfunctions of the
covariant Laplacian, which is a 
hermitian operator with a strictly positive spectrum,
\ba 
-\Delta_\mu^2[U]f^{(n)}(x)&=&
\sum_{\mu}\left[2f^{(n)}(x)-U_\mu(x)f^{(n)}(x+\hat{\mu})-U^\dag_\mu(x-\hat{\mu})
f^{(n)}(x-\hat{\mu})\right]\nn\\
&=&\lambda^n f^{(n)}(x),\quad \lambda^n>0. 
\label{lev}
\ea
Its eigenvectors have the desired transformation property $f^{(n)g}(x)=g(x)f^{(n)}(x)$.
They provide a unique mapping $U\rightarrow f[U]$ except when eigenvalues are degenerate
or $|f|=0$. In practical simulations the probability of generating such
configurations is essentially zero \cite{vi}.
These properties have been used previously for gauge fixing
without Gribov copies \cite{vw} and to construct blockspins for the derivation of
effective theories \cite{hh}.
The eigenvectors are non-local in the sense that they depend on all link variables.
In order to maintain the transfer matrix formalism the $f(x)$ have to be local in time.
This is achieved by considering the spatial Laplacian $\Delta_i^2[U_i]$ in \eq{(\ref{lev}), which
then is defined in every timeslice individually and independent of $U_0$.

The eigenvectors are used to construct an $N\times N$ matrix
$\Omega(x)\in SU(N)$ following \cite{vw}. Since \eq (\ref{lev}) only determines them up to
a phase, this leaves a remaining freedom in $\Omega(x)$.
In the case of $SU(2)$, to every eigenvector $f$ there is a degenerate second one
given by its charge conjugate $\ii\tau^2f^*$.
In order to have a smooth field the solution to the lowest eigenvalue is chosen in practice.
These may now be combined into 
\be\label{om1}
\Omega(x)\equiv\frac{1}{|f^{(0)}(x)|}\left(
\begin{array}{*{2}{c}{c}}
f^{(0)}_1(x) & -f^{(0)*}_2(x)\\
f^{(0)}_2(x) & f^{(0)*}_1(x)
\end{array}\right) \quad \in SU(2).
\ee
The whole matrix satisfies the Laplace equation, $-\Delta_i^2[U]\Omega=\lambda^0\Omega$,
and so does $\Omega h$, where $h$ may be any global
$SU(2)$ matrix. 
For $SU(3)$ there is no degeneracy of the eigenvalues in general. In this case
one solves for the three lowest eigenvectors to construct the  matrix 
$\Omega = (f^{(0)},f^{(1)},f^{(2)})$,
which is then determined
up to a factor $h={\rm diag}(\exp(\ii \omega_1),\exp(\ii \omega_2),\exp(\ii \omega_3)),
\sum_i\omega_i=0$. (For an alternative construction and 
a numerical implementation, see \cite{afo}).
The remaining indeterminacy $h$ may be different for every timeslice.
This is summarized by the transformation behaviour
\be \label{otrafo}
\Omega^g(x)=g(x)\Omega(x)h^\dag (t),
\ee
where $h(t)$ is free.

We can now define composite link and gluon fields
\ba \label{cl}
V_\mu(x)&=&\Omega^\dag(x)U_\mu(x)\Omega(x+\hat{\mu}),\\
A_\mu(x)&=&\frac{\ii}{2g}\left[V_\mu(x)-V^\dag_\mu(x)
-\frac{1}{N}\tr\left(V_\mu(x)-V^\dag_\mu(x)\right)\right],\nn
\ea
both transforming as
$O_i^g(x)=h(t)O_i(x)h^\dag(t)$, whereas $V_0^g(x)=h(t)V_0(x)h^\dag(t+1)$.
Hence the $A_i(x)$ are gauge invariant under spatial transformations $g(\bfx)$,
but transform under time-dependent rotations corresponding to the residual symmetry
of the spatial Laplacian. Note that the functions $\Omega$ and hence the composite
gauge fields have no definite $Z_N$ symmetry and couple to all sectors.

With the abbreviation 
\be
V_0(\bfx;t_1,t_2)=\prod_{t'=t_1}^{t_2-1}V_0(\bfx,t')
\ee
for a temporal Wilson line,
we can now construct the gauge invariant Euclidean operator
\be \label{gluestring}
O[U]=\tr \left[A_i(\bfx,0)V_0(\bfy;0,t)
A_i(\bfx,t)V_0^\dag(\bfz;0,t)\right].
\ee
It represents a correlator in $t$
of the composite field $A_i$, where Wilson lines are inserted
to ensure full gauge invariance. Note however, that these may be placed at any $\bfy,\bfz$.
For the particular choice $\bfx=\bfy=\bfz$, the Wilson lines
merge into one adjoint representation line
$V_{0,ab}^A(\bfx;0,t)=
\tr[T^aV_0(\bfx;0,t)T^bV_0^\dag(\bfx;0,t)]$,
\be \la{lump}
O[U]= \frac{1}{2} A^a_i(\bfx,0)V_{0,ab}^A(\bfx;0,t)A^b_i(\bfx,t) .
\ee
\eq (\ref{lump})
represents a correlator describing a gluon bound to an adjoint source,
in analogy to the ``gluelump'' operators \cite{mic}. In Appendix \ref{gs} 
the corresponding continuum expression is derived by coupling the gluon field
to explicit fields for the static sources, which are then integrated 
out analytically.

In \cite{me} the transfer matrix formalism has been employed 
to convert \eq (\ref{gluestring})
into a quantum mechanical trace,
\be \label{L0ham}
\langle O[U]\rangle  = 
 Z^{-1}\trq\left\{\htm^{N_t-t}\hat{A}_{i,\alpha\beta}(\bfx)\;
\htm_{L0}^t \;
\hat{A}_{i,\alpha\beta}(\bfx)\right\},
\ee
where we have defined a modified transfer matrix $\htm_{L0}$ by the 
`Laplacian temporal' gauge in which $V_0(x)=1$.
It has been proved that $\htm_{L0}$ has the same spectrum as $\htm_{0}$ \cite{me}.
Initially we had only known that $\htm_0$ is
invariant under time independent gauge transformations. Now we 
have the stronger result that $\htm_0$ and $\htm_{L0}$ have the same spectrum, 
even though they are related by a time dependent gauge transformation. 
This finding is the lattice analogue to a continuum quantization of
the Schr\"odinger functional: equivalent results are obtained in temporal 
and any other gauge that is invariant under spatial rotations \cite{rossi}.

Hence the expectation value of our operator 
decays exponentially with eigenvalues of the Hamiltonian
$\hat{H}_0$,
\be \label{finalen}
\lim_{N_t\rightarrow \infty} \langle O[U] \rangle = \sum_{n}
|\langle 0|\hat{A}_{i,\alpha\beta}(\bfx)
|n\rangle |^2 \exp -(E_n-E_0)t\;.
\ee
\eqs (\ref{gluestring}),(\ref{L0ham}) and (\ref{finalen}) are the desired 
gluon analogue to \eq (\ref{qmpot}).

\subsection{Existence of a continuum limit}

When the continuum limit is approached,
the exponents extracted from 
\eq(\ref{finalen}) diverge because of the self-energy
contributions of the temporal Wilson lines, \fig \ref{div}. 
In order to retain a finite continuum limit, the operator
has to be modified such that no divergent mass renormalization 
is present.

%%%%%%%%%%%%%%%%%%%%%%%%%%%%%%%%%%%%%%%%%% FIGURE
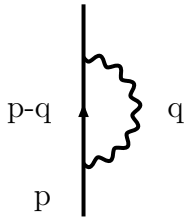
\begin{figure}[tbh]

%\vspace*{-0.9cm}

\begin{center}

\begin{picture}(150,120)(0,0)

\SetWidth{1.5}
\ArrowLine(62,20)(62,100)
\PhotonArc(62,60)(20,-90,90){1.5}{8}

\Text(50,26)[r]{p}
\Text(50,59)[r]{p-q}
\Text(94,59)[l]{q}

%\ArrowLine(122,20)(122,100)
%\ArrowLine(118,100)(118,20)
%\PhotonArc(122,60)(20,-90,90){1.5}{8}

%\Text(14,0)[l]{(a)}
%\Text(114,0)[l]{(b)}

\end{picture}
\vspace*{-0.4cm}
\end{center}
\caption[]{\label{div}\it
The Wilson line self-energy. \eq (\ref{tav}) enforces $p=p-q=0$.}

\vspace*{0.5cm}

\end{figure}
%%%%%%%%%%%%%%%%%%%%%%%%%%%%%%%%%%%%%%%%%%%%%%%%%%%

This can be achieved by observing that the transformation behaviour of $V_0(\bfx,t)$ in \eq
(\ref{cl}) is independent of the spatial coordinates. In the construction of the operator
\eq (\ref{gluestring}), instead of $V_0$ we may then
use its timeslice average
\be \label{tav}
\tilde{V_0}(t)=\sum_{\bfx}V_0(\bfx,t)/||\sum_{\bfx}V_0(\bfx,t)||,
\ee
which has been projected back into the group.
The timeslice average corresponds to the Fourier transform of $V_0$ with zero momentum.
If this is done
in every timeslice, the sources represented by the $\tilde{V}_0$ cannot emit
a gluon at one $t$ and reabsorb it at some later $t$ as in Fig.\ref{div}.
Hence, the mass renormalization of the static source is switched off, and the sources
remain classical external fields.
The presence of fields $\tilde{V}_0$
then merely accounts for the transformation behaviour, but has
no effect on the gauge field energies
measured by the modified operator
\be \label{ofinal}
\langle O[U]\rangle=\left\langle \tr \left[A_i(\bfx,0)\tilde{V}_0(0,t)
A_i(\bfx,t)\tilde{V}_0^\dag(0,t)\right]\right\rangle,
\ee
which has a spectral decomposition as in \eq (\ref{finalen}).
The energies extracted from the expectation values of
\eqs (\ref{gluestring}) and (\ref{ofinal}) should then differ
by a cut-off dependent shift due to the selfenergy contribution \fig \ref{div}.
A non-perturbative parton mass for the gluon may now be defined by the gap 
between the lowest excitation
energy and the vacuum,
\be
m_A\equiv E_1-E_0.
\ee
By Fourier transformation of the spectral representation \eq (\ref{finalen})
it follows trivially that the energy eigenvalues appear as 
poles in momentum space. It
 depends on the behaviour of the higher energy levels 
whether in the infinite volume and contiuum limits the lowest 
eigenvalue remains an isolated pole or
turns into a cut.

\subsection{Non-uniqueness of $\Omega$ and the relation to gauge fixing}

How do the above results depend on the particular construction of $\Omega$?
Clearly, any $\Omega[U]\in SU(N)$ local in time and
satisfying the transformation law \eq (\ref{otrafo}),
permits construction of a gauge invariant observable \eq (\ref{ofinal}). From the 
spectral representation it follows that all such observables fall off with the same
spectrum, while $\Omega$ only enters the matrix elements and thus influences
the overlap of the operator with the eigenstates. 

Of course, the construction of the composite link variable \eq (\ref{cl})
may also be viewed
as fixing Laplacian gauge on each timeslice \cite{vw}.
This gauge is incomplete, with
a global factor $h$ remaining unfixed between time-slices.
It can be completed by imposing
the further condition $\tilde{V}_0(t)=1$, thus fixing $h(t)$.
In this particular gauge the operator
\eq (\ref{ofinal}) is equivalent to a gauge fixed gluon propagator,
falling off exponentially with eigenvalues of the transfer matrix.
Since the spectrum is unaffected by the particular construction of $\Omega$, this statement
holds for all gauges employing a unique $\Omega[U]$
local in time.
For example, the standard Coulomb gauge is defined by an $\Omega(x)$ that 
minimizes the functional
\be \label{coul}
R[U]=\sum_{x,i}\left[1-\frac{1}{N}\tr U^{\Omega^\dag}_i(x)\right]
\ee 
in every timeslice. $\Omega(x)$
has the desired properties and a residual freedom $h(t)\in SU(N)$,
to be fixed in the same way. A recent implementation is in~\cite{ckp}.
(Of course, this gauge condition has the problem that it does not determine
$\Omega$ uniquely \cite{rev}). 
Hence, any gauge fixing done in this manner is equivalent to coupling the
flux to sources and averaging over all their positions.
In momentum space this non-perturbative result translates into what is also found
in perturbation theory: the pole of a propagator is gauge independent, whereas the
residue, viz.~the matrix element, is not. 
On the other hand, Landau gauge is non-local in time,
no positive transfer matrix is defined and non-perturbatively 
it is not guaranteed that it probes
the same spectrum.

\section{\label{num} Numerical tests for SU(2) in 2+1 dimensions}

This section is devoted to numerically test the operator constructed in the last
section. This first exploratory study is done for SU(2) 
in 2+1 dimensions, for its significantly lower numerical cost and fast continuum approach
permitting to extract conclusive continuum results.
In this case the coupling constant $g^2$ has dimension of mass and provides the 
scale in which all physical results are expressed.

\subsection{\label{whiggs} W-boson in a Higgs model}

It is useful to first test the operator \eq (\ref{ofinal}) in a situation where the result
is known and controlled by perturbation theory. 
We therefore consider a SU(2) Higgs model with
scalar $2\times 2$ matrix fields $\Phi(x)$ 
in the fundamental representation and the continuum action 
\be\label{l3d}
S = \int d^3x \; \tr \left[{1\over 2}F_{ij}F_{ij} +
(D_i\Phi)^\dag D_i\Phi + m_0^2 \Phi^\dag \Phi
+ 2 \lambda (\Phi^\dag \Phi)^2 \right] \, .
\ee
Its physical properties are determined
by the two dimensionless parameters 
\be
x=\frac{\lambda}{g^2},\quad y=\frac{m_0^2}{g^4}\;.
\ee
The lattice action and parameters as well as their relation to the continuum ones
are given in Appendix \ref{lata}.
%%%%%%%%%%%%%%%%%%%%%%%%%%%%%%%%% FIGURE
\begin{figure}[ht]
\vspace*{0.5cm}
\centerline{\epsfxsize=6cm\hspace*{0cm}\epsfbox{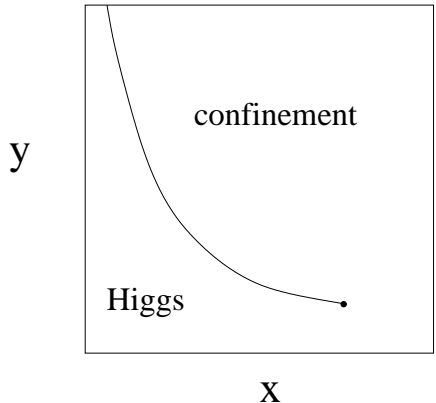}}

\caption[a]{{\em
Schematic phase diagram of the Higgs model \eq (\ref{l3d}). Pure gauge theory is
reached in the limit $x,y\rightarrow \infty$.}}
\la{pd}
\end{figure}
%%%%%%%%%%%%%%%%%%%%%%%%%%%%%%%%%%%

The phase diagram of the theory has been determined non-perturbatively in the continuum limit
and is shown schematically in \fig \ref{pd}: Higgs and confinement regions
are separated by a line of first order phase transitions, which ends in
a critical point \cite{hels}. 
The spectrum in the respective regions as well as 
its continuous connection through the crossover
region has been studied extensively \cite{us1,us2}. 

In the Higgs region, the theory has 
a physical vector boson
carrying the quantum numbers of the gluon, coupling to the
gauge invariant composite operator 
\ba 
W_i^a(x)&=&\tr\left(T^a\Phi^\dag(x) D_i\Phi(x)\right) \hspace*{1.5cm} \mbox{(continuum)},\nn\\
W_i^a(x)&=&\tr\left(T^a\Phi^\dag(x) U_i(x)\Phi(x+\hat{\i})\right) \quad \mbox{(lattice)}.
\la{wbos}
\ea
Its mass is determined by the decay of the corresponding correlation function 
\be \label{wcor}
\langle W_i^a(\bfx,0)W_i^a(\bfx,t)\rangle \sim \ex^{-M_V t}.
\ee
$M_V$ is an eigenvalue of the Hamiltonian acting on 
the gauge invariant subspace ${\cal H}$, and the W-boson is an asymptotic particle state.
On the other hand, in perturbation theory one fixes a gauge and computes
the vector boson mass from the fall-off of the gauge field propagator,
\be
\langle A_i^a(\bfx,0)A_i^a(\bfx,t)\rangle \sim \ex^{-m_A t}\ .
\ee
In momentum space the mass corresponds to the renormalized pole of the propagator.
At a finite order in perturbation theory one finds $m_A= M_V$.
With our operator constructed in the last section, \eq (\ref{ofinal}), we now have 
a lattice implementable
gauge field correlator without any reference to the scalar fields. We non-perturbatively
know it to decay exponentially, where
$m_A$ is an eigenvalue of the Hamiltonian acting on the entire Hilbert space ${\cal H}_0$.

\fig \ref{comp} shows a comparison of the ground state mass
as obtained from the two correlation functions, 
measured at one point in the Higgs phase, $x=0.0239,y=-0.02$, 
for different lattice spacings.
Lattice sizes were chosen large enough for the masses to be free of finite volume effects,
based on the results of \cite{us1}.
Within statistical errors, the two operators yield identical results.
The mass extracted from the new operator indeed extrapolates to a physical
continuum limit, as promised in Sec.~\ref{nlgl}.
As expected for an object with dimension
$mass^2$, the lowest eigenvalue of the spatial covariant Laplacian diverges strongly
in the continuum limit. For the correlation function this does not matter because 
the eigenvalue itself does not enter the definition
of the correlation function. 

%%%%%%%%%%%%%%%%%%%%%%%%%%%%%%%%% FIGURE
\begin{figure}[ht]

\centerline{\epsfysize=6.0cm\hspace*{0cm}\epsfbox{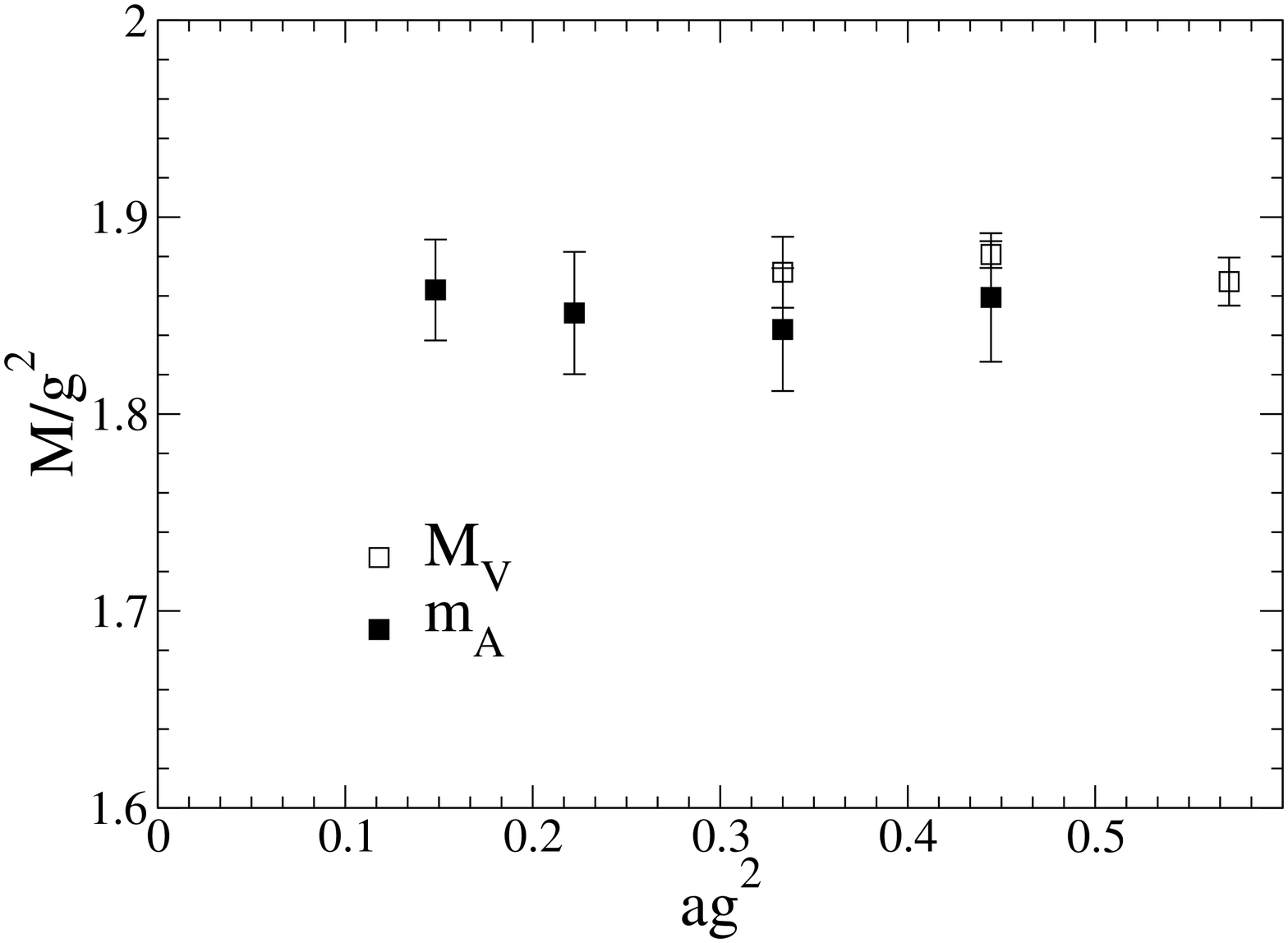}
\epsfysize=6.0cm\hspace*{0.6cm}\epsfbox{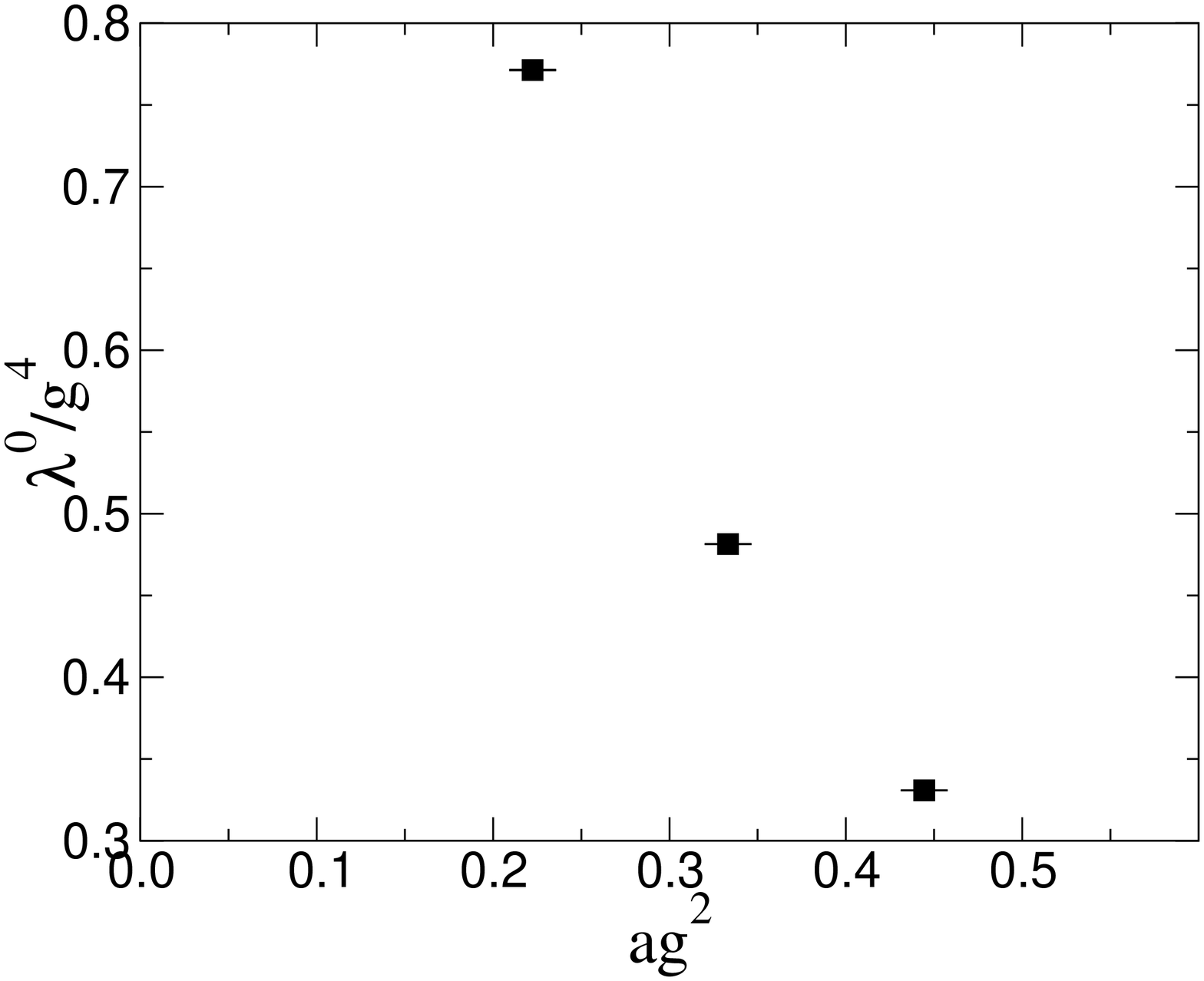}}

\caption[a]{{\em 
{\rm Left:} 
The W-boson mass in a 3d Higgs phase, computed from the gauge invariant composite operator
$W_i^a[\phi,U]$, \eq (\ref{wbos}), 
and the gauge field propagator, \eq (\ref{ofinal}).
The data for $M_V$ are from \cite{us1}.
{\rm Right:} The lowest eigenvalue of the lattice Laplacian, \eq (\ref{lev}).}}
\la{comp}
\end{figure}
%%%%%%%%%%%%%%%%%%%%%%%%%%%%%%%%%%%%

We then conclude that the gauge field propagator has a non-perturbative pole
in the Higgs region of the model, and within our accuracy we find $m_A=M_V$
also non-perturbatively.
While this may not be surprising, we stress that it is a
non-trivial result. What has been known is that the perturbatively defined pole
in the field propagator gives a good approximation of the physical W-mass in
suitable gauges and when the perturbative Higgs expectation value is 
large compared to quantum fluctuations \cite{fro,wb}, i.e.~when perturbation theory
is reliable.
However, it has not been clear whether this pole also exists non-perturbatively,
since that requires a
non-perturbative definition for $m_A$.
Moreover, it is important to realize that, because of the residual 
gauge freedom in the composite fields, \eq (\ref{cl}), we have
\be
\left \langle  A^a_i[U]\;
W^a_i[\phi,U] \right\rangle =0,
\ee 
i.e.~the operators are orthogonal,
and so are the corresponding eigenstates of the Hamiltonian. This is
yet another reflection of the fact that the eigenstates of the Hamiltonian 
$\hat{H}_0$ contributing to the correlator \eq (\ref{ofinal})
are in a different sector of the Hilbert space than the eigenstates
contributing to the correlator \eq (\ref{wcor}).
Consequently, $m_A,M_V$ are really different
eigenvalues of $\hat{H}_0$ which are degenerate in a Higgs dynamics only. 

The analytic connectedness of the phase diagram implies that
there is a one to one mapping of the
the entire spectrum of the Hamiltonian $\hat{H}_0$ 
between Higgs and confinement regions of the phase diagram \cite{fs}. 
For example, the W-boson state described by the operator \eq (\ref{wbos}), 
being an eigenstate of the projected transfer matrix $\htm$,
becomes a vector meson bound state in the confinement region \cite{us1,us2}.
As we shall see, $m_A\neq M_V$ in a confining regime, as is also found in
resummed perturbation theory \cite{wb,bp} and in lattice Landau gauge \cite{kar1}.
The parameters can be continuously tuned to $x,y\rightarrow \infty$, where the 
scalar fields are infinitely heavy and decouple to leave us with a pure gauge theory. 
All meson states including $M_V$ disappear from the (finite) spectrum in this limit.
The question then is what happens to the pure gauge quantity $m_A$.

\subsection{\label{ym1} Yang Mills theory}

In addition to a gluon, one may also
construct other operators from the composite links \eq (\ref{cl}).
In particular, simply taking the trace,
\be
O_V(x)=\tr(V_i(x)),
\ee
produces an operator coupling to $0^{++}$ states. It is fully invariant under
gauge transformations, the residual freedom $h$ in the definition of the composite
link discussed in Sec.~\ref{nlgl}
drops out under the trace. Hence, the correlation $\langle \op_V(x) \op_V(y) \rangle $
falls off with the spectrum of the projected Hamiltonian $\hat{H}$, and
one expects to be able to extract the $0^{++}$ glueball mass from it.

\fig \ref{mvl} shows the masses in lattice units extracted from the correlators in
the $0^{++}$ and $1^{--}$ channels, respectively, plotted against the spatial length of the 
lattice, $L$. 
At intermediate and large distances we see a linear rise with $L$, which is reminiscent
of torelons, i.e.~flux loop states winding around the lattice \cite{mtor}. 
The well-known $0^{++}$ torelon couples to Polyakov loops $P_i^{(L)}(x)=\prod_{n=0}^{L-1}\,
U_i(x+n{{\hat{\imath}}})$, winding around a spatial direction of the lattice. 
Its mass extracted from the corresponding correlation function is related
to the string tension in a finite volume,
\be
%\label{string}
  \left\langle \tr P_i^{(L)}(\bfx,0)\;
   \tr P_i^{(L)}(\bfx,t)
                  \right\rangle \simeq \ex^{-am_{P^{(L)}}t},\quad
   am_{P^{(L)}}=a^2\sigma_L L.
\ee
Torelons are physical states in the gauge invariant sector in a finite volume,
but become infinitely heavy and decouple when the infinite volume limit is taken.
As \fig \ref{mvl} demonstrates, the two operators $O_V, \tr P^{(L)}$ give 
identical results and we conclude that $O_V$ projects onto the torelon.

%%%%%%%%%%%%%%%%%%%%%%%%%%%%%%%%% FIGURE
\begin{figure}[th]
\vspace*{1cm}
\centerline{\epsfxsize=10cm\hspace*{0cm}\epsfbox{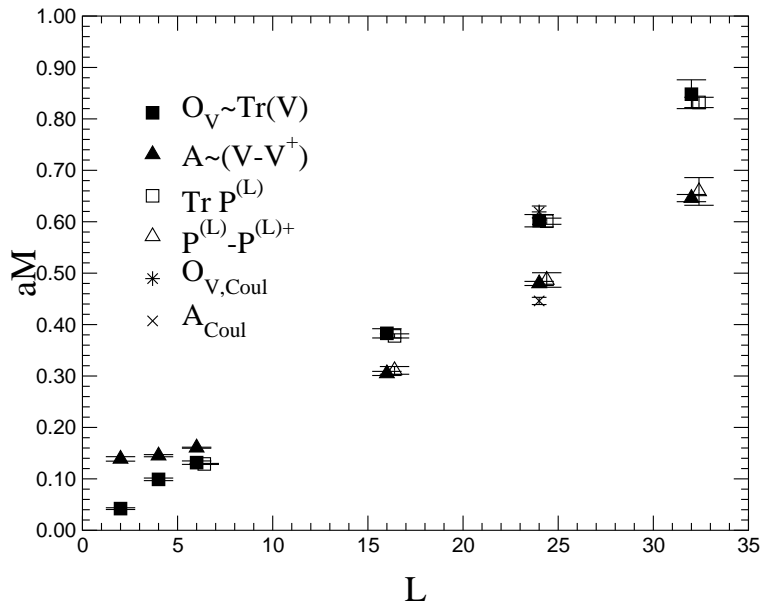}}

\caption[a]{{\em
Effective masses of $1^{--}$ and $0^{++}$ operators
using the composite link $V[U]$ and Polyakov loops, as described in the text.
The simulation is in YM-theory at $\beta=9$ with spatial lattice size $L$.}}
\la{mvl}
\end{figure}
%%%%%%%%%%%%%%%%%%%%%%%%%%%%%%%%%%%%

This suggests to interpret the lower $1^{--}$ state as 
torelonic state as well. In order to confirm this, we compare with the
traceless part of the Polyakov loop. The latter is not gauge invariant, but if we replace
the spatial link by a spatial Polykov loop, $U_i\rightarrow P_i^{(L)}$ in \eq (\ref{cl}),
the correlator \eq (\ref{ofinal}) is again gauge invariant.
The resulting masses are also
shown in the plot and found to be identical to the ones
from the original correlator.
Hence we conclude that we have found a new eigenstate of the Hamiltonian for
a pure gauge theory in a finite volume, corresponding
to a torelon in an adjoint state. 
Because it transforms non-trivially, 
it is not an asymptotic state like the standard torelon, 
but rather an eigenstate
of $\hat{H}_0$. 

For $L=24$, we have also calculated the correlators with an $\Omega[U]$ computed from the
Coulomb gauge fixing condition \eq (\ref{coul}). As \fig \ref{mvl} shows, the results
are, in accordance with our theoretical considerations, equivalent to those
obtained from the Laplacian procedure.
In particular, since the $\Omega$ emerging in Coulomb gauge is also non-local,
it projects on the same winding states. 
The small deviation observed in the $1^{--}$ mass 
might be related to the influence of Gribov copies.
  
\fig \ref{mvl} is essentially unchanged
when computed in the confining regime of the Higgs model rather than
in the pure gauge theory. This is in agreement with previous findings that 
the confining gauge sector of the Higgs model is almost insensitive
to the presence of the scalar fields \cite{us1,us2}. 
It confirms that
the gluon correlator \eq (\ref{ofinal}) does not project on 
the vector meson with mass 
$M_V$, but rather on a pure gauge quantity.

\subsection{Localized vs.~torelon states}

Does the adjoint torelon state have anything to do with the gluon?
For finite $L$ this is not the case, 
just like the $0^{++}$ torelon is not related to the glueball. 
The analytic connectedness of the
phase diagram holds for the theory in infinite volume. This implies that 
the infinite volume pole in the gluon 
correlator found in the Higgs regime cannot disappear from the
spectrum when the parameters are changed to the confining regime. 

The reason for seeing only torelonic states is a projection problem of
the maximally non-local operators:
the functional $\Omega[U]$ depends on all links in
the spatial volume. In a Higgs regime this is of no import because 
colour is screened and 
the links far from each other do not communicate. In a confining regime on the
other hand, 
adjacent links form flux tubes stretching over the whole spatial lattice.
An operator constructed from $\Omega$ must have a large overlap with 
such non-local objects, and a very small overlap with localized states.
Put in another way, in the composite gauge field the link to be correlated, 
$U_i(x)$, is ``drowned'' by the dynamics of all other links contributing through
$\Omega[U]$, a situation that worsens as the spatial volume is increased.
For example, at some $L$
the $0^{++}$ torelon becomes heavier than the lightest glueball, which then should dominate
the correlation function of $O_V$. At the lattice spacing chosen in \fig \ref{mvl}, 
the lightest glueball mass
is $aM_G=0.755(7)$ \cite{tep}. At $L=32$ the torelon is heavier, yet \fig \ref{mvl}
demonstrates that $O_V$ is blind to the localized glueball state.

This finding is corroborated by the fact that the problem is absent 
when correlations other than those
of pure gauge operators are studied. For example, propagators of an adjoint scalar field
$\Phi(x)$ coupled to the gauge fields have been studied in the Coulomb gauge in
\cite{ckp}. No significant finite volume dependence
was found for the scalar propagator. While the
corresponding composite 
operator $\Omega^\dag\Phi(x)\Omega(x)$ also employs a non-local function
$\Omega[U]$, it is local in the scalar field to be correlated.

We then conclude that pure gauge operators constructed from the composite 
links $V_i$ in a confining dynamics 
have very weak overlap with localized states, and in their current form
are not very useful to extract infinite volume physics.
Nevertheless, a new eigenvalue of the Hamiltonian in the sector
with quantum numbers of the gluon has been found.
Just like the $0^{++}$ torelon is related to the potential
of static charges in a singlet state, the adjoint piece of the flux tube should
be related to the potential of static charges in a triplet 
state (octet for SU(3)) \cite{ls,nad}.
This question is beyond the scope of the present paper and will be pursued
elsewhere. However, having extracted the energy of a flux tube in an adjoint state,
note that it extrapolates to a non-zero value $\sim 0.32g^2$ 
for $L\rightarrow 0$. 
In our Hamiltonian considerations we have interpreted the gluon mass as the minimal
energy of precisely such a piece of flux.
However, one would expect this value to be affected by finite volume effects,
as one side of the lattice has been shrunk to zero.

\section{\label{stati} The gluon mass and static mesons}

In order to circumvent the projection problem encountered in the previous
section, we now seek to compute the energy eigenvalue of interest
from local observables coupled to sources. 
Rather than averaging over all their positions, the contribution of the sources
will be computed separately and subtracted.
This approach will also provide a relation
between the gluon propagator in temporal gauge
and the ratio of static meson correlators, 
leading to a physical interpretation of the
gluon mass in terms of level splittings. 

For this purpose we go back to the expressions of a gluon coupled to
external sources in the continuum, \eq(\ref{gs}\ref{statv}) 
from Appendix A, and on the lattice, \eq (\ref{lump}).
For a non-perturbative evaluation,
we want to avoid the non-local functions 
used in the definition of the composite lattice gauge field.
This may be achieved by observing that the
energies governing the decay of a correlator are completely determined by the
Hamiltonian and the quantum numbers of the operators.
Instead of discretizing the gauge field,
we may then equally well employ a local higher 
dimension operator sharing the same quantum numbers, such as the 
adjoint part of the
linear combination of plaquettes in the spatial plane transverse to the
Wilson line,
\be
\op_{DF}(x)=P_{i,j}(x)+P_{-i,j}(x)
+P_{-j,-i}(x)+P_{-j,i}(x)-{\rm h.c.}
\ee
In the continuum limit,
the operator approaches the covariant derivative of the field strength,
$D_jF_{ij}(x)$, and 
couples to $J^{PC}=1^{--}$ states.
We are then considering the two-point function 
\be \label{gll}
\langle \op^a(\bfx,0)U_{0,ab}^A(\bfx;0,t)\op^b(\bfx,t)\rangle \sim \ex^{-M_{\op}t},
\ee
with an adjoint temporal Wilson line $U_{0,ab}^A(\bfx;0,t)$.
%\tr[T^aU_0(\bfx;0,t)T^bU_0^\dag(\bfx;0,t)]$.
It has the general form of 
a gluelump correlator,
describing a bound state of dynamical glue and a 
static adjoint source \cite{mic}. Hence we are looking for the mass of
the lightest $1^{--}$, or ``electric'', gluelump.

Because we are now working with undressed temporal links rather
than with composite ones, the zero momentum projection \eq (\ref{tav})
cannot be applied, and the observable   
still contains the divergent contribution
from the adjoint Wilson line and its self-energy.
We now need to compute this contribution separately and
subtract it in order to get a finite result in the continuum limit.
The Wilson line can be made gauge invariant by closing it through the
periodic boundary in the $t$-direction, in which case it 
becomes an adjoint Polyakov loop,
whose exponential decay with the temporal lattice size $N_t$ 
we need to compute.
On the other hand, the Polyakov loop couples to 
adjoint states in all $J^{PC}$ sectors, and asymptotically its exponential
decay is the same as that of \eq (\ref{gll}),
where $\op(x)$ is the operator leading to the smallest coefficient $M_{\op}$.
Hence, we are looking for the mass of the lightest gluelump, which is well 
known \cite{mic} to be the ``magnetic'' one described by the clover field,
\be
\op_F(x)=P_{i,j}(x)
+P_{j,-i}(x)+P_{-i,-j}(x)+P_{-j,i}(x)-{\rm h.c.}
\ee
In the continuum, this operator approaches the field strength
$F_{ij}(x)$, coupling to $0^{--}$ and $1^{+-}$ states
in 2+1 and 3+1 dimensions, respectively.

The lowest eigenvalue of $\hat{H}_0$ determining the asymptotic decay of the
gluon correlation function \eq (\ref{ofinal}) should then be the same
as the gluelump mass splitting 
\be
m_A=M_{DF}-M_F\;.
\ee
Note that the gluelump states may be viewed as
the zero distance limit of static potentials with the string between
the sources being in an excited state specified 
by appropriate quantum numbers \cite{mic}.
This is in accordance with our interpretation of the gluon mass
as a short distance property of potentials put forward in the Hamiltonian
strong coupling analysis, Sec.~\ref{ham}.
The calculation of gluelump splittings 
involves only pure gauge operators local in space and is
straightforward to evaluate on a lattice, with well controllable infinite volume
and continuum limits.

\subsection{\label{gl3d} Numerical result for SU(2) in 2+1 dimensions}

Gluelumps in three dimensions have been simulated previously in different
contexts \cite{poul,ml}. Here we are specifically interested
in a continuum extrapolation of the mass splitting $M_{DF}-M_F$.
The numerical results of a calculation in SU(2) pure gauge theory
close to the continuum are displayed in \fig \ref{extra}.
The choice of lattice sizes follows the results of \cite{ml}, where 
it has been explicitly
checked that the lowest gluelump is free of finite volume effects.
%%%%%%%%%%%%%%%%%%%%%%%%%%%%%%%%% FIGURE
\begin{figure}[ht]

\vspace*{1cm}
\centerline{\epsfxsize=10cm\hspace*{0cm}\epsfbox{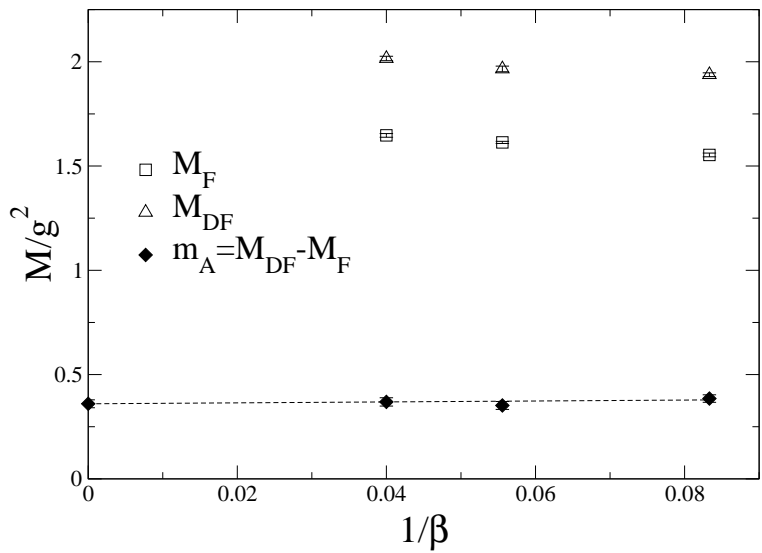}}
\caption[a]{{\em
Masses for the electric ($M_{DF}$) and magnetic ($M_F$) gluelumps, as well as
their mass splitting. Calculations have been performed for $\beta=12,18,25$
on lattices with $L=30,48,64$, respectively.
The mass splitting is extrapolated to the continuum.}}
\la{extra}
\end{figure}
%%%%%%%%%%%%%%%%%%%%%%%%%%%%%%%%%%%
The figure displays the onset of the weak logarithmic divergence in the
gluelump masses, as well as the expected scaling for their mass difference.
A linear extrapolation of the data in $1/\beta$ to the continuum then
gives the final result for the SU(2) gluon mass in 2+1 dimensions
\be
m_A=0.360(19)g^2,\qquad \chi^2/{\rm dof}=0.47.
\ee
Comparing with the lightest scalar
glueball, $M_G=1.584(17)g^2$ \cite{tep}, one finds $M_G/m_A= 4.4(3)$.

\subsection{The magnetic mass}

Besides being an interesting test case, the gauge theory in 2+1 dimensions
plays an important role in four-dimensional physics at finite temperature.
Representing the Matsubara zero mode sector of the latter, it constitutes
the effective theory describing all static physics at asymptotically high
temperatures. 
A long standing problem of thermal field theories are the severe infrared 
divergencies encountered in perturbation theory.
In particular, loops of magnetic gauge fields $A_i$ exhibit divergences
on the three-dimensional scale $g^2\sim g_{4d}^2T$.
These may be cured by dynamical generation of
a ``magnetic mass'' which, however, is entirely non-perturbative and
receives contributions from all orders in a loop expansion \cite{ir}.
Here we have shown that such a magnetic mass can indeed be defined non-pertubatively,
and that for hot SU(2) gauge theory its asymptotic high temperature value
is $m_{A}=0.36(2)g_{4d}^2T$.

\begin{table}[ht]
\begin{center}
\begin{tabular}{|c|cc|cc|}
\hline
               & ref. && $m_A/g^2$& \\ \hline
1-loop gap eq. & \cite{an} &   & 0.38 & \\
               & \cite{bp,jp}& & 0.28 &\\
               & \cite{co}  & & 0.25 &\\
2-loop gap eq. & \cite{eb}   & & 0.34 &\\ \hline
lattice MAG & \cite{kar} & & 0.51(6)&\\ \hline
\end{tabular}
\end{center}
\caption[]{\label{mmag}
\it Comparison of magnetic mass values from
gap equations and gauge fixed lattice simulations.}
\end{table}

In the past, gauge invariant resummation
schemes have been designed to compute the pole of the gluon propagator
self-consistently in three dimensions \cite{bp},\cite{an}-\cite{eb}.
In a Hamiltonian
analysis of the three dimensional gauge theory a gauge invariant composite gluon
variable has been constructed, which in the weak and strong coupling limits yields
a gluon mass gap as the lowest eigenvalue of the Hamiltonian \cite{nair}.
One may now compare the results of these approaches collected in Table
\ref{mmag} with the full answer and get
an estimate for the quality of the approximations involved. 
It appears that the resummations lead to reasonable answers. In particular the
two-loop result exhibits convergence towards the result of the last section.
Note that lattice simulations employing maximal abelian 
gauge fixing provide numbers that are
somewhat off our result, underlining the difficulty of obtaining stability in 
such calculations. In Landau gauge the correlator becomes negative at
large distances, and it appears difficult to extract a mass \cite{ckp}. 

\subsection{Numerical result for SU(3) in 3+1 dimensions}

The considerations to relate the gluon mass in pure gauge
theory to the splitting of magnetic and electric gluelumps 
are directly applicable to SU(3) Yang-Mills theory in 3+1 dimensions.
The only difference is that the lowest magnetic gluelump and the
corresponding operator in four dimensions
has $J^{PC}=1^{+-}$.
The mass splitting between these lowest states has been computed in \cite{glmic},
which quotes for the continuum extrapolation
\be \label{su3m}
m_A=M_{DF}-M_{F}=368(7){\rm MeV}.
\ee
Comparing with the lightest scalar glueball, $M_G[0^{++}]=1730(130)$ MeV \cite{mp},
we then have $M_G/m_A=4.7(5)$, which is strikingly similar to
our 2+1 dimensional calculation in SU(2). This is not unexpected, given the
close resemblance of other properties
of pure gauge theories in 2+1 and 3+1 dimensions.
Together with the ratio $M_G/m_A$, Table \ref{2vs3} collects results for
the mass of the lightest scalar glueball and the string breaking distance
for the flux tube between adjoint sources, 
suggesting that the confining low energy
physics is very similar in these theories.

Similar to the situation in three dimensions, a larger value $m_A\approx 600$ MeV 
has recently been estimated from the gluon propagator in 
Laplacian gauge \cite{lapg}. However, while this gauge is free from Gribov copies,
no positive transfer matrix exists and hence the functional form of the decay is
not exactly known.

\begin{table}[ht]
\caption{\em Comparison of some continuum extrapolated
pure gauge quantities between 2+1 and 3+1 dimensions.
\label{2vs3}
}
\begin{center}
\begin{tabular}{|c| r@{.}l r@{.}l|}
\hline
&\multicolumn{2}{c}{2+1}&\multicolumn{2}{c|}{3+1}\\
\hline
SU(2), $M_G[0^{++}]/\sqrt{\sigma}$& 4&72(4) \cite{tep} & 3&74(12) \cite{tep1} \\
SU(3), $M_G[0^{++}]/\sqrt{\sigma}$& 4&33(4) \cite{tep} & 3&64(9) \cite{tep1}\\
SU(2), $M_G r^{adj}_{break}$   & 10&3(1.5) \cite{adj1} & $\sim$ 9&7 \cite{adj2}\\
SU(2), $M_G/m_A$               & 4&4(3)  & \noner \\
SU(3), $M_G/m_A$               & \none  & 4&7(5) \cite{mp,glmic} \\
\hline
\end{tabular}
\end{center}
\end{table}

\section{\label{qcd} Mass splittings between heavy mesons}

In this section we ask what happens in the vicinity of the static limit,
when heavy but dynamical matter fields are present, and whether the
static limit can be attained smoothly.
One might hope that the gluon
mass is then observable in the spectrum of sufficiently heavy mesons.
Two candidate splittings appear as possibilities.

First, hybrid meson states are, in the non-relativistic approximation,
bound states in gluonic excitations of the static potential.
These potentials have been calculated \cite{hpot}, and in the limit
of small distances
appear to smoothly connect to the gluelump spectrum \cite{glmic}.
In view of this the splitting of the lowest $1^{--}$ and $1^{+-}$ 
hybrid mesons, computed in the appropriate potentials, should
be close to the gluon mass. However, these states have so far been
elusive experimentally.

The second possibility is a splitting between a hybrid and a 
non-hybrid state. Since spin effects are suppressed, let us consider 
QCD with scalar quarks as
specified in Appendix \ref{gs}.
We are interested in the correlator of the vector meson 
\be
V_i(x)=\im\left(\phi^\dag(x)U_i\phi(x+\i)\right),\quad J^{PC}=1^{--},
\ee
but now we wish to keep the bare scalar mass $m_0$ finite
rather than integrating
over the scalar fields analytically.
The main difference to the static case concerns 
the subtraction of the contribution
of the scalar fields. Away from the static limit they will not sit 
at the same point in an adjoint state, but rather form spatially extended
singlet mesons, the lightest of which couples to
\be
S(x)=\re\left(\phi^\dag(x)\phi(x)\right),\quad J^{PC}=0^{++}.
\ee
Comparing the operator content of $S,V$, one may think of $V$ as a
hybrid meson.
The energy difference between vector and scalar mesons is accounted for by
three different contributions: {\it i)} the total spin
$\vec{J}=\vec{L}_\phi+\vec{J_g}$, consisting of angular momentum of the scalars
and the gluon spin; excitation of the {\it ii)} scalar and {\it iii)}
gluon fields in higher quantum states.
In the limit $m_0\rightarrow \infty$ the scalars become static
and hence their angular momentum is
switched off. Furthermore, they are ``quenched'' to be external fields
and cannot be excited into higher quantum states. 
The remaining contributions to the mass splitting are the gluon spin
and excitations of the gluon field.
In this case the 
vector meson becomes spin-exotic, i.e.~its quantum numbers cannot be
accounted for without gluon degrees of freedom.
Thus, the mass difference between vector and scalar meson in the static limit
is probing a gluonic excitation with the quantum
number of the gluon. This suggests that the gluon mass should be approached
arbitrarily accurately by the limit
\be \label{def}
m_A= \lim_{m_0\rightarrow\infty}[M_V-M_S].
\ee
If the limit exists, the last equation may even serve as an alternative
definition of the gluon mass. However, formally
it is not clear whether a smooth 
limit exists.
Calculating the scalar correlator in
the static limit in the continuum using 
\eqs (\ref{gs}\ref{2step}),(\ref{gs}\ref{slim}), one obtains
\be
\langle S(x)S(y)\rangle
\stackrel{m_0\rightarrow\infty}{\longrightarrow}
g_0^2(x,y) \left\langle \tr\left(U_0(x,y)
U^\dag_0(x,y)\right)\right\rangle \;,
\ee
which has the same form on the lattice.
The temporal Wilson lines combine
to a unit matrix, hence the desired cancellation of 
their renormalization with
that of the $V$ correlator does
not work and, strictly speaking, the formal static limit of $M_V-M_S$ does not 
have a finite continuum limit.
The reason is that static sources in a singlet state annihilate
when they are at the same point. The only way for them to exist at zero separation is
to combine their colour into an $(N^2-1)$-plet state, which we have used in the
previous sections to calculate the static limit.
However, this problem is absent for
arbitrarily large but finite $m_0$, and one may ask whether the value for the gluon mass
can be smoothly approached by $M_V-M_S$ for arbitrarily large $m_0$.
As we shall see in the next section, this is indeed the case in 2+1 
dimensions.

\subsection{\label{shiggs} Heavy mesons for SU(2) in 2+1 dimensions}

We now turn to a test of this proposition 
in the SU(2) lattice gauge theory in 2+1 dimensions. The scalar QCD action 
\eq (\ref{gs}\ref{gphi}) can then be rewritten in terms of scalar $2\times 2$ matrix fields
$\Phi(x)$ and discretized as given in
Appendix \ref{lata}.
Results of detailed simulations for varying scalar mass $m_0$ are shown 
in \fig \ref{mhiggs}, which displays the scalar and vector mesons,
together with the lightest scalar glueball and its first excitation.

Note that scalar glueballs and mesons
are indistinguishable by quantum numbers. To tell one from another a mixing analysis
following \cite{us1,us2} has to be performed. Quite generally, 
the glueballs and scalar mesons show very little mixing even when
they are close in mass, and thus are easy to identify. 
An example is shown in \fig \ref{mhiggs} (right), where the matrix elements between the
$0^{++}$ mass eigenstates $\Psi_n$ and the operators used in the simulation 
$\op_i$ are displayed for $m_0=1.0g^2$. 
Here the operators are ``smeared'' 
versions of pure gauge and scalar operators, 
$\op_i\in\{G=\tr F_{ij}^2,S=\phi^\dag\phi\}$. (For details of the 
mixing analysis in a Higgs model,
cf. \cite{us1}). Beginning with the lowest one, the three $0^{++}$ states are easily
identified as glueball, scalar bound state and glueball, based on their operator content,
and the situation is similar for the other parameter values.

The dotted error bands in \fig \ref{mhiggs} (left)
give the location of the lightest scalar glueballs in the pure gauge theory \cite{tep}. 
It has already
been reported for various scalar gauge models in 2+1 dimensions that the glueball spectrum 
deviates 
only at the percent level from that in the pure gauge theory \cite{us1,us2,hp}.
Indeed the figure shows that already for bare masses $m_0\gsim 2g^2$ the glueball spectrum
assumes its pure gauge values to high numerical accuracy,
indicating that the scalars have largely decoupled as dynamical fields.

%%%%%%%%%%%%%%%%%%%%%%%%%%%%%%%%% FIGURE
\begin{figure}[ht]

\vspace*{1cm}
\centerline{\epsfxsize=8cm\hspace*{0cm}\epsfbox{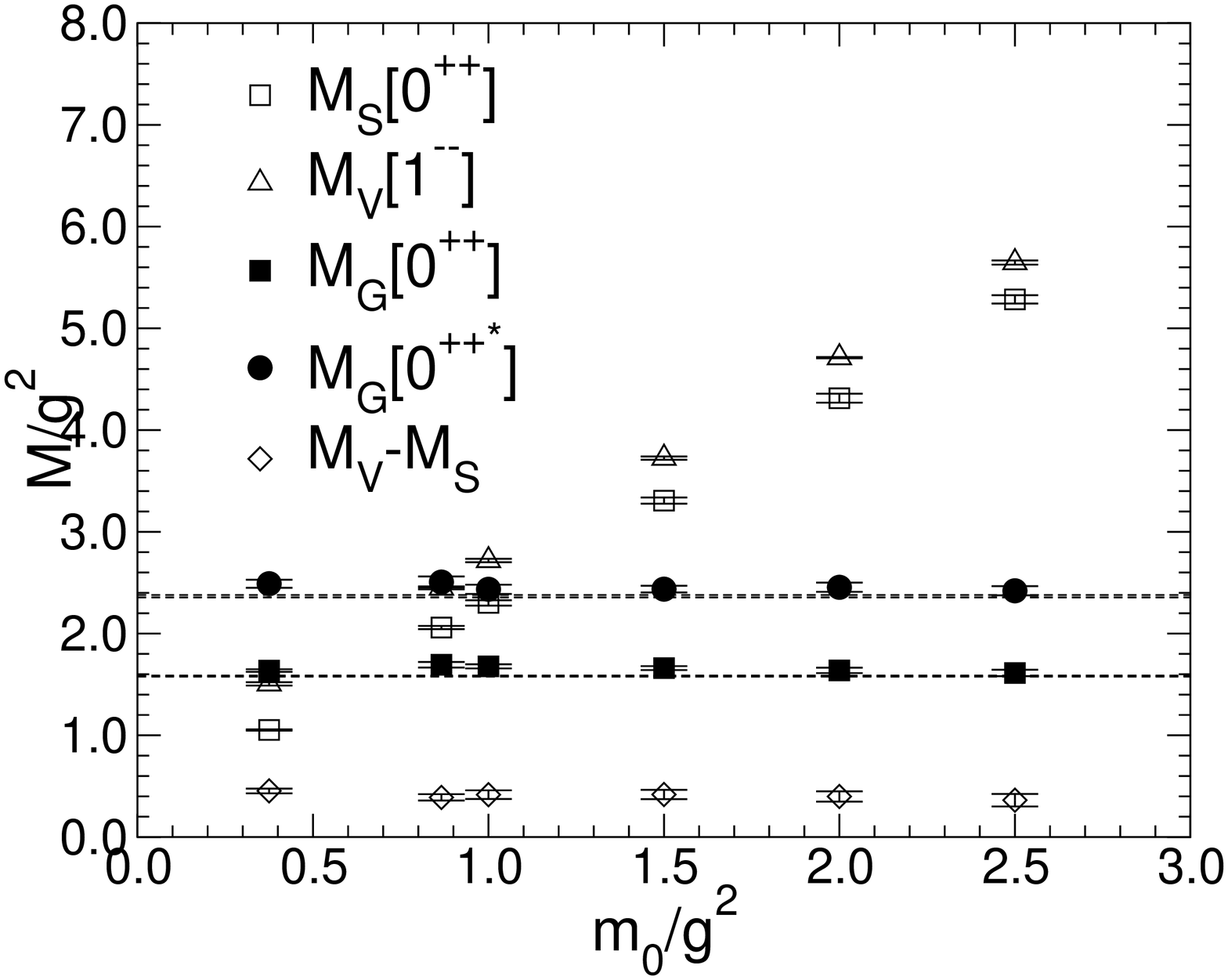}
\epsfxsize=8cm\hspace*{0.5cm}\epsfbox{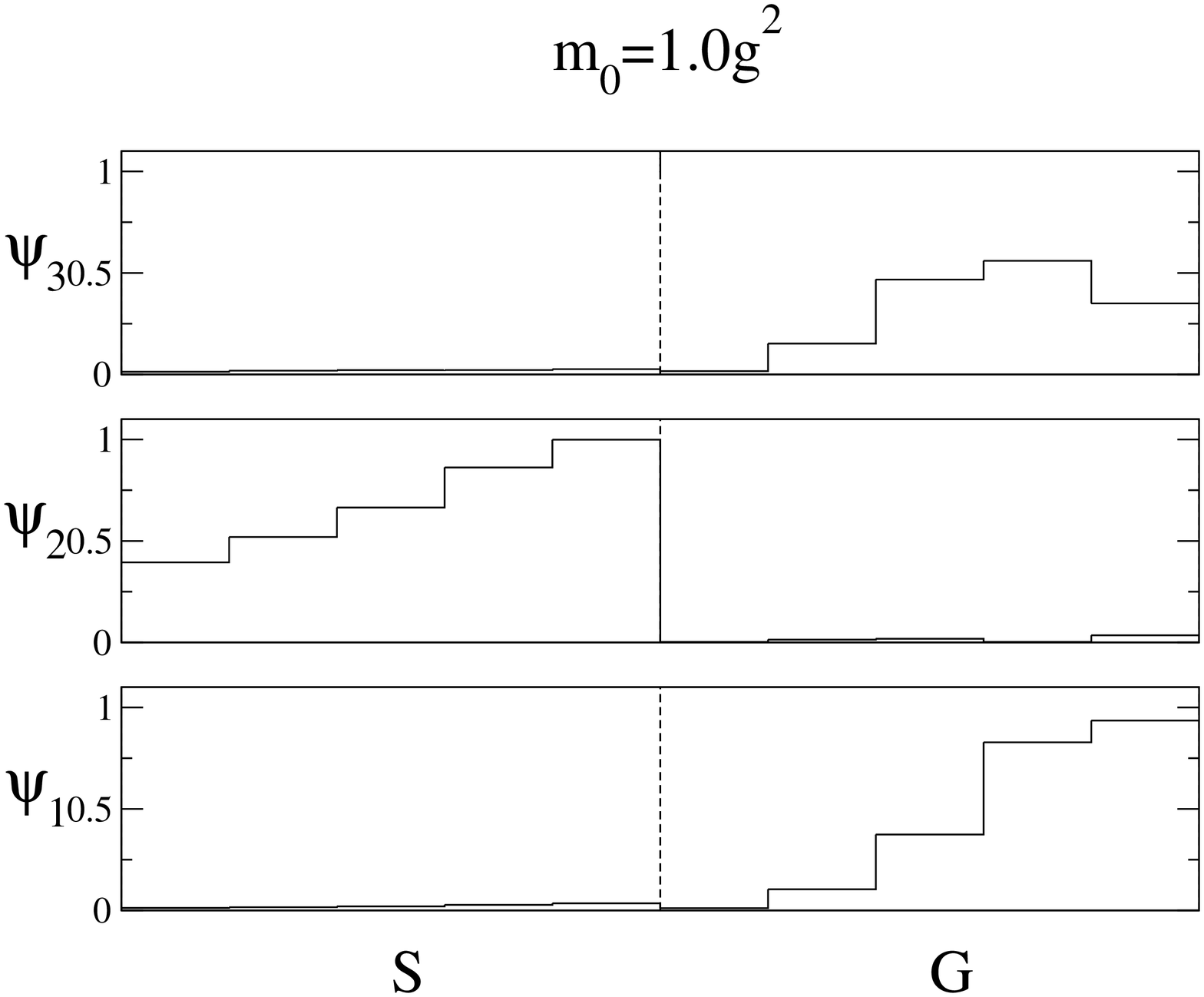}}
\caption[a]{{\em
{\rm Left:} The lowest states in 3d scalar QCD, simulated at $\beta=18$ on $48^3$
lattices. 
$M_G$ denotes scalar glueballs, $M_{S,V}$
scalar and vector mesons.\\
{\rm Right:} Overlap $\langle\Psi_i|\op_j|0\rangle$ 
between the three lowest states $\Psi_i$ in the $0^{++}$ channel and
blocked operators, $\op_j\in\{S,G\}$.}}
\la{mhiggs}
\end{figure}
%%%%%%%%%%%%%%%%%%%%%%%%%%%%%%%%%%%
Also shown is the mass splitting between vector and scalar mesons, which is only slightly
diminishing over the range of $m_0$ studied here. 
This is in accordance with the fact that we are close to the pure gauge limit.
Note however, that in three dimensions the Coulomb part of the static potential is
logarithmic. In a logarithmic potential, the level splittings of bound states of scalars 
calculated from a non-relativistic Schr\"odinger equation are also independent
of the constituent mass \cite{qr}, so that constancy of the splitting alone in this
case cannot indicate the vicinity of the pure gauge limit. 

According to the prescription \eq (\ref{def}), 
extracting the gluon mass requires an extrapolation
of the splitting to infinite scalar masses.
However, for heavier scalar fields the numerical errors grow rapidly and an
accurate determination of the mass difference is increasingly difficult.
This is well known from heavy quark physics and 
can in principle be cured by NRQCD methods \cite{nrqcd}.
Taking the largest value of $m_0$ in \fig \ref{mhiggs} one reads off
$m_A=0.37(6)$. This is in excellent agreement with the static limit 
calculated in Sec.~\ref{gl3d}. We then conclude that in scalar QCD in 
2+1 dimensions the gluon mass is observable to good accuracy in the mass splitting
between vector and scalar mesons, which smoothly connects to the static limit.

\subsection{Heavy quarkonia in QCD}

Similar to the pure gauge physics in Table \ref{2vs3}, 
a close resemblance is known between the SU(2) Higgs models in 2+1 and 
3+1 dimensions. Both phase diagrams are analytically connected,
with mass spectra of Higgs and W-bosons or bound states of scalars and glueballs
in the Higgs and confinement regions, respectively (for a review and references, see
\cite{hi4d}).
Moreover, in the symmetric region the confinement and screening of fundamental
charges as well as the mixing properties of the gluonic flux with 
meson states are the same in both models \cite{sbr}, 
and one might expect the results of the 
previous section to carry over. 
However, there
is also a significant difference concerning the short range Coulomb part
of the static potential, which is now $\sim r^{-1}$.
While a calculation analogous to the one in Sec.~\ref{shiggs} is beyond the scope
of the current paper, it is intriguing to speculate 
about the situation in QCD, which provides us with heavy quarks
to probe the gluon dynamics.

We then consider the mass splitting of
heavy quarkonia instead of scalar mesons. Away from the static limit 
the situation is complicated by the quark spins,
the total meson spin now being $\vec{J}=\vec{L}_q+\vec{J}_g+\vec{S}_q$.
Spin independent meson masses analogous to the case of heavy scalar fields
(and static sources) are obtained by averaging over the quark spin multipletts.
In the standard spectroscopic notation (based on the quark model) the total meson spin is 
written as $\vec{J}=\vec{L}+\vec{S}_q$.
Hence $\vec{L}=\vec{L}_q+\vec{J}_g$, which is now playing the role of 
$\vec{J}=\vec{L}_\phi+\vec{J}_g$ in the scalar case.
Thus, based on a comparison of quantum numbers, the splitting $M_V-M_S$ 
in the theory with heavy scalars might correspond to
the well known spin averaged mass 
splitting $\Delta_{1P-1S}$, as put forward in a preliminary 
account of this work \cite{berl}.
Using current experimental numbers \cite{prev} one finds 
$\Delta_{1P-1S}(\bar{c}c)=418.5$ MeV and
$\Delta_{1P-1S}(\bar{b}b)=416$ MeV. 

However, it is not clear that these numbers are tied to the gluon mass.
The fact that they are practically identical and quark mass independent cannot be taken
as an indication of the proximity of the static limit, but is also accounted for
by the effectively logarithmic non-relativistic 
bound state potential at the length scales relevant for these
quarkonia\footnote{I thank W.~Buchm\"uller for pointing this out to me.} \cite{qr,log}.
Increasing the quark mass pushes the states into the $\sim r^{-1}$ 
Coulomb region of the potential, in which mass splittings
scale linearly with the quark mass. Hence the splittings of 
orbitally excited quarkonia do not have a static limit. On the other hand, this statement
concerns only states within the non-relativistic 
quark model. In QCD there should be additional states,
the hybrids, in which gluonic excitations account for the same quantum numbers, and
one would expect the static limit of the splitting to exist.
In general mixing analyses are necessary in order to unambiguously 
identify the nature of an observed state. 
A four-dimensional analogue of the
calculation in Sec.~\ref{shiggs} could clarify 
this question.

\section{\label{con} Conclusions}

It has been demonstrated that an unambiguous, non-perturbative definition of a gluon mass
is possible in terms of an eigenvalue of the Kogut-Susskind Hamiltonian of lattice
gauge theory, which can be interpreted as the energy of a gluon coupled to 
static sources. This eigenvalue can be computed in several ways. 
It dictates the asymptotic exponential decay of non-local pure gauge observables which, 
in a particular gauge, reduce to the gauge field propagator. Using the eigenvectors
of the covariant Laplacian for its construction, these observables are free 
of Gribov copies and strictly positive, with effective masses being upper bounds
to the ground state. In practice these operators are 
numerically feasible in Higgs regimes,
but not in confining dynamics, where their non-local nature results in an almost
exclusive projection onto torelonic states.

However, the same eigenvalue also governs
the asymptotic exponential decay of ratios of static gluelumps,
in which gluon fields are bound to adjoint sources.
This establishes a non-perturbative relation between the gluon mass and 
manifestly
gauge invariant observables. These can be easily computed using local operators,
with well controlled infinite volume and continuum limits. 
Furthermore, in three dimensions 
the mass splitting of vector and scalar mesons approaches the same
quantity in the static limit. In four dimensions this might be modified
due to the different nature of the Coulomb potential.

Viewing the lattice field theory as a statistical system, it is clear that this
gluonic mass scale 
plays a fundamental role in colour dynamics: $1/m_A$ is the largest
correlation length in the pure gauge system, providing an infrared cut-off for 
virtual states as well as setting a scale for the screening of colour interactions.
For thermal physics, in particular, 
this should allow a non-abelian definition of Debye screening 
in complete analogy to QED. 
Moreover, the fact that $m_A$ in three dimensions is finite implies
a non-zero magnetic mass which we have computed for SU(2), 
and thus the screening of colour magnetic fields in a plasma.
A physical interpretation of the zero temperature gluon mass
requires more work to investigate the static limit of meson mass splittings.

We then conclude that a general gauge invariant description of colour dynamics 
in terms of partonic degrees of freedom should
be possible, with numerous interesting questions to be answered. First, it would be
very desirable to solve the projection problem of the non-local gluon propagators and
numerically confirm the relations between their exponential decay and the static
mesons established here.
An obvious generalization would be to apply the same techniques to quark propagators,
which might offer an alternative approach to non-perturbative quark mass renormalization.
Similar methods applied to Polyakov loops should also allow
to give a gauge invariant meaning to the finite temperature 
static potential with sources in an octet state.
Finally, one may hope to get a new handle on some proposed confinement mechanisms.
For example, a gluon getting massive dynamically is in acccord with what is
expected in a picture of the vacuum as a dual superconductor.

\begin{appendix}

\section{\label{gs} Continuum pure gauge theory with static sources}

Let us introduce a complex scalar $N$-plet $\phi_\alpha(x)$, $\alpha=1,...N$,
coupled to the gauge fields, with the action
\be \label{gphi}
S_{\phi}[A,\phi]=\int d^3x\;\left\{(D_\mu\phi)^*(x)D_\mu\phi(x) + m_0^2\phi^*(x)\phi(x)\right\}.
\ee
Together with the pure gauge action,
the theory corresponds to QCD with scalar ``quarks''.
In the limit $m_0\rightarrow \infty$ the terms $|D_i\phi|^2$
are suppressed and decouple from the above action.
In this case the scalar fields propagate in time only, describing static sources
coupling to colour electric flux of the gauge fields.
The scalar propagator satisfying
$\left(-D_0^*D_0+m_0^2\right)G(x,y)=\delta(x-y)$
is known exactly:
\be
G(x,y)=g_0(x,y)\delta(\bfx,\bfy)U_0(x,y),
\ee
where $g_0(x,y)$ is the free scalar propagator without background field
and $U_0(x,y)$ the temporal Wilson line from $x$ to $y$.

We are interested in observables of the type
$\op(x_1,x_2)=\phi^\dag(x_1)M(x_1,x_2)\phi(x_2)$,
where $M(x_1,x_2)$ is a pure gauge matrix operator containing Wilson lines and/or
covariant derivatives.
Doing the Gaussian integral
over the scalars one obtains for the correlation function ($x_i=(\bfx_i,0),y_i=(\bfx_i,t)$)
\be \label{2step}
\langle \op(x_1,x_2) \op^\dag(y_1,y_2)\rangle=
Z^{-1}\int DA \;\;\bar{\op}[A]\;\ex^{-S_{YM}[A]},
\ee
with
\ba \label{obs}
Z&=&\int DA\; Z_\phi \;\ex^{-S_{YM}[A]}, \quad
Z_\phi=[\det(-D^2+m_0^2)]^{-1},\nn\\
\bar{\op}[A]&=& Z_\phi^{-1}
\int D\phi D\phi^*\; \op(x_1,x_2) \op^\dag(y_1,y_2)\;\ex^{-S_\phi[A,\phi]}.
\ea
In the limit $m_0\rightarrow \infty$ the scalar determinant becomes a constant
and $\bar{\op}[A]$ represents a pure gauge quantity,
\be \label{slim}
\bar{\op}[A]=g_0(x_1,y_1)g_0(x_2,y_2)\;\tr[M(x_1,x_2)U_0(x_2,y_2)M^\dag(y_1,y_2)U_0(y_1,x_1)].
\ee
A prominent example is the string of colour flux ending on
scalar charges separated by $r=|\bfx_1-\bfx_2|$, which is described by the
operator $M(x_1,x_2)=U(x_1,x_2)$.
In the limit $m_0\rightarrow\infty$ the correlator goes over into the Wilson loop,
\be
\langle \op(x_1,x_2) \op(y_1,y_2)\rangle
\stackrel{m_0\rightarrow\infty}{\longrightarrow}
 g^2_0(t)\langle W(r,t) \rangle ,
\ee
decaying exponentially with the energy of the string in the presence of sources.

In the same way as the energy of a flux tube,
the gluon energy can be probed by coupling it to scalar sources.
In making contact to the lattice strong coupling gluon wave function \eq (\ref{psitot}),
we consider the simplest
gauge invariant operator with $J^{PC}$ quantum numbers of the gluon,
\be \label{mops}
%S(x)&=&Re\left(\phi^\dag(x)C(x,x)\phi(x)\right),\quad J^{PC}=0^{++},\nn\\
%S(x)&=&Re\left(\phi^\dag(x)\phi(x)\right),\quad J^{PC}=0^{++},\nn\\
V_i(x)=\im\left(\phi^\dag(x)D_i\phi(x)\right),\quad J^{PC}=1^{--}.
\ee
The connection to the gluon propagator is established by
integrating over the scalars analytically.
Using \eqs (\ref{2step}),(\ref{slim}),
one finds for the correlator
\be \label{statv}
\langle V_i(x)V_i(y)\rangle
\stackrel{m_0\rightarrow\infty}{\longrightarrow}
-g_0^2(x,y) \left\langle \tr\left[D_i(x)U_0(x,y)D_i(y)U^\dag_0(x,y)\right]
\right\rangle .
\ee
Since the expression is manifestly gauge invariant, it may be evaluated in any gauge.
Choosing the temporal gauge, $A_0(x)=0$, we have $U_0(x,y)=1$ and 
$D_i(x)U_0(x,y)\rightarrow -2igA_i(x)$,
The right hand
side of \eq (\ref{statv}) is thus 
identical to the gluon propagator in temporal gauge times
a free scalar propagator,
\be
\lim_{m_0\rightarrow \infty}
\langle V_i(x)V_i(y)\rangle\left.\right|_{A_0=0}=
4g^2g_0^2(x,y)\left.\langle\tr A_i(x)A_i(y)\rangle\right|_{A_0=0}.
\ee
Note that the pure gauge operator in \eq (\ref{statv}) is
simply the continuum version of
\eq (\ref{gluestring}).
It is then clear that its exponential decay is determined
by the spectrum of the continuum Hamiltonian in temporal gauge.
However, in the continuum, $D_i(x)$
is an operator defined at a point, which allows its correlator
to be written down without the introduction of non-local functions of the gauge field.

\section{\label{lata} Lattice actions}

In this appendix we give the lattice action and parameters used for the simulations 
in the various parts of this paper. 
\\
\noindent
{\bf SU(2) Higgs model, Sec.~\ref{whiggs}:}
The lattice action for the fundamental representation Higgs 
model in 2+1 dimensions with $2\times 2$ matrix field $\Phi(x)$, cf.~\eq (\ref{l3d}), 
may be defined as
\ba \label{actlat}
S[U,\Phi]&=&\beta\sum_p\left(1-\frac{1}{2}\tr U_p\right)
+\sum_x\Bigg\{-\beta_H\sum_{i=1}^3\frac{1}{2}
\tr\Big(\Phi^{\dagger}(x)U_i(x) \Phi(x+\hat{i})\Big)  \nn\\
&&+\frac{1}{2}\tr\Big(\Phi^{\dagger}(x)\Phi(x)\Big)
+\beta_R\left[\frac{1}{2}\tr\Big(\Phi^{\dagger}(x)\Phi(x)\Big)-1\right]^2
\Bigg\}.
\ea
The relation between the parameters of the continuum action \eq (\ref{l3d})
and those of the lattice action is at the two loop level \cite{lai}
\ba \label{lcp1}
\beta&=&\frac{4}{ag^2},\nn \\
\beta_R&=&x \frac{\beta_H^2}{\beta},\nn \\
y&=&\frac{\beta^2}{8}\left(\frac{1}{\beta_H}-3
-\frac{2\beta_H}{\beta}x
\right)+\frac{3\Sigma\beta}{32\pi}\left(1+
4x\right)\nn \\
& &+\frac{1}{16\pi^2}\left[\left(\frac{51}{16}+9x
-12x^2\right)
\left(\ln\frac{3\beta}{2}+\zeta\right)+5.0+5.2x\right]
\; ,\label{lcp3}
\ea
with the numerical constants $\Sigma=3.17591$ and $\zeta=0.09$.
Due to superrenormalizability, \eq (\ref{lcp3}) are exact up to discretization errors.
The continuum limit is at one point in
the lattice phase diagram, $\beta_G \rightarrow\infty, \beta_H=1/3, \beta_R=0$.
\\
\noindent
{\bf Yang-Mills limit, Secs.~\ref{ym1}, \ref{gl3d}:}
The pure gauge action is seen to be a limit of the above obtained by taking
$\beta_H=\beta_R=0$. From \eq (\ref{lcp3}) it is evident that this corresponds
to sending the continuum parameters $x,y\rightarrow \infty$. 
\\
\noindent
{\bf Scalar QCD, Sec.~\ref{shiggs}:}
Finally, the Gaussian scalar action simulated in Sec.~\ref{shiggs} is obtained
by taking $\beta_R=0$.

The algorithm used to perform the Monte Carlo simulation using the
action in \eq (\ref{actlat}) is the same as in~\cite{us1,us2}.
The gauge variables are updated
by a combination of heatbath and over-relaxation steps
according to \cite{fab,ken}, while the scalar degrees of freedom
are updated combining heatbath and reflection steps
as described in \cite{bunk}.
The ratio of the different updating steps is suitably tuned
such as to minimize autocorrelations.
In the simulations we gathered typically between 5 000 and 20 000
measurements taken after such combinations of updating sweeps.

In order to increase the overlap of operators with the low energy states
to be measured, standard ``smearing'' or ``blocking'' techniques 
have been applied to gauge
and scalar fields. The correlation matrix between operators of different 
smearing levels is measured and diagonalized by a variational calculation
allowing to extract ground and excited states in a given quantum number channel.
For details and references see \cite{us1}.

\end{appendix}

\section*{Acknowledgements}
\noindent
It is a pleasure to acknowledge numerous valuable discussions with 
O.~B\"ar, W.~Buchm\"uller, P.~de Forcrand, J.~Negele, U.-J.~Wiese
and H.~Wittig.
I thank P.~de Forcrand for computer code to invert the Laplacian as
well as to fix Coulomb gauge. The simulations in this work have been
performed on a NEC/SX32 at the HLRS Stuttgart.

\end{document}